\documentclass[12pt]{iopart}
\usepackage{iopams}
\usepackage{longtable}
\usepackage{amssymb}
\usepackage{graphicx}
\usepackage{dcolumn}

\begin{document}

\title{Proton-Neutron Pairs in Heavy Deformed Nuclei}

\author{Dennis Bonatsos, I.E. Assimakis, Andriana Martinou}

\address{Institute of Nuclear and Particle Physics, National Centre for Scientific Research Demokritos,
GR-15310 Aghia Paraskevi, Attiki, Greece}

\ead{bonat@inp.demokritos.gr}

\begin{abstract}
The microscopic justification of the emergence of SU(3) symmetry in heavy nuclei remains an interesting problem. 
In the past, the pseudo-SU(3) approach has been used, with considerable success. 
Recent results seem to suggest that the key for understanding the emergence of SU(3) symmetry lies in the properties of the proton-neutron interaction, namely in the formation of (S=1, T=0) p-n pairs in heavy nuclei, especially when the numbers of valence protons and valence neutrons are nearly equal. Although this idea has been around for many years, since the introduction of the Federman-Pittel mechanism, it is only recently that information about the p-n interaction could be obtained from nuclear masses, which become available from modern facilities. Based on this information, a new coupling scheme for heavy deformed nuclei has been suggested and is under development. 

\end{abstract}


\maketitle

\section{Introduction}

In the present work, some questions in nuclear structure open for several decades are addressed. 

1) Why is the SU(3) symmetry present in heavy nuclei \cite{IA}, while it is known that the SU(3) symmetry 
of the 3D harmonic oscillator \cite{Elliott} is destroyed beyond the sd shell by the strong spin-orbit interaction \cite{Mayer}? 

2) In the framework of the pseudo-SU(3) scheme \cite{pseudo1,pseudo2}, an approximate SU(3) symmetry is obtained 
for the normal parity levels of each major nuclear shell, while the opposite parity level is kept separate. 
On the other hand, according to the Federman--Pittel mechanism \cite{FP1,FP2}, spin-orbit partner orbitals 
of protons and neutrons of opposite parity are mainly responsible for nuclear deformation. How are these 
two approaches reconciled? 

The main points of the present work are the following.

1) We are going to argue that deformation in heavy deformed nuclei is due to proton--neutron pairs of Nilsson orbitals 
of the type $\Delta K[\Delta N \Delta n_z \Delta \Lambda]=0[110]$ \cite{Cakirli}, 
where $K$ and $\Lambda$ are the projections of the total and orbital angular momenta on
the $z$-axis ($K$=$\Lambda$$\pm$1/2), while \textit{N} ($N$=${n_x}$+${n_y}$+${n_z}$) is the oscillator quantum number,
and $n_z$ is the number of quanta in the $z$-direction (the deformation
axis), in agreement with the Federman--Pittel mechanism.

2) In addition, we are going to argue that an approximate SU(3) symmetry different from the pseudo-SU(3) one is present, 
in which orbitals of both parities are treated on equal footing \cite{Bonatsos}. 
 
\section{Filling of shells}

From the standard Nilsson diagrams one can form a table with the order in which 
Nilsson orbitals of protons and Nilsson orbitals for neutrons are filled.
Although the order is modified with increasing deformation, it is modified 
in similar ways, as one can see in Fig. 4 of Ref. \cite{Bonatsos}. 

 The relevant lists for protons in the 28-50 major shell and neutrons  in the 50-82
shell are given in  the left hand side of Table 1. The standard notation $K[N n_z \Lambda]$ is used.
The following remarks apply:


\begin{table} [b]
\caption{Nilsson orbitals appearing in different major shells of the nuclear shell model.}\smallskip
\begin{small}\centering
\begin{tabular*}{\textwidth}{@{\extracolsep{\fill}}rrrr|rrrr}
\hline  \noalign {\smallskip}
  & p:28-50  &   & n:50-82  &  & p:50-82 & & n:82-126 \\ 
\hline\noalign
{\smallskip}
 1 & 1/2[321] &  1 & 1/2[431] &  1 & 1/2[431] & 1 & 1/2[541] \\
 2 & 3/2[312] &  3 & 1/2[420] &  2 & 3/2[422] & 3 & 1/2[530] \\
 3 & 1/2[310] &  2 & 3/2[422] &  3 & 1/2[420] & 2 & 3/2[532] \\
 4 & 3/2[301] &  7 & 1/2[550] &  4 & 1/2[550] & 4 & 1/2[660] \\
 5 & 5/2[303] &  8 & 3/2[541] &  5 & 3/2[541] & 5 & 3/2[651] \\
 6 & 1/2[301] &  4 & 3/2[411] &  6 & 5/2[532] & 6 & 5/2[642] \\
 7 & 1/2[440] &  9 & 5/2[532] &  7 & 5/2[413] & 8 & 3/2[521] \\
 8 & 3/2[431] &  5 & 5/2[413] &  8 & 3/2[411] & 7 & 5/2[523] \\
 9 & 5/2[422] &  6 & 1/2[411] &  9 & 7/2[523] & 9 & 7/2[633] \\
10 & 7/2[413] &    & 5/2[402] & 10 & 1/2[411] &10 & 1/2[521] \\
11 & 9/2[404] &    & 7/2[404] & 11 & 7/2[404] &12 & 5/2[512] \\
   &          & 10 & 7/2[523] & 12 & 5/2[402] &11 & 7/2[514] \\
   &          &    & 1/2[400] & 13 & 9/2[514] &   & 7/2[503] \\
   &          & 11 & 9/2[514] & 14 &11/2[505] &   & 9/2[505] \\
   &          &    & 3/2[402] & 15 & 3/2[402] &13 & 9/2[624] \\
   &          &    &11/2[505] & 16 & 1/2[400] &16 & 1/2[510] \\
   &          &    &          &    &          &15 & 3/2[512] \\
   &          &    &          &    &          &14 &11/2[615] \\
   &          &    &          &    &          &   &13/2[606] \\
   &          &    &          &    &          &   & 3/2[501] \\
   &          &    &          &    &          &   & 5/2[503] \\
   &          &    &          &    &          &   & 1/2[501] \\
   
\hline
\end{tabular*}
\end{small}
\end{table}

1) For each proton orbital there is a corresponding $\Delta K[\Delta N \Delta n_z \Delta \Lambda]=0[110]$ neutron orbital.
We shall call these pairs ``0[110] partner orbitals''. 
To exhibit the correspondence, the proton orbitals are labelled by their order, 1 to 11, starting from the one lying lowest in energy.
Then the number preceding the neutron orbital indicates its 0[110] ``partner'' proton orbital. 
 
2) We remark that the first 9 orbitals occupied by protons and neutrons are the same. 
They are not filled in exactly the same order, but once they are filled for a given nucleus, 
the order is not important any more. 

3) Since in the neutron shell there are 5 orbitals more than in the proton shell, 
5 neutron orbitals have to stay alone. All these orbitals are characterized by $n_z=0$. 
All unpaired orbitals appear in the upper half of the shell. 

A similar picture is seen for protons in the 50-82 shell and neutrons in the 82-126 shell, 
shown in the right hand side of Table 1, as well as for protons in the 82-126 shell and neutrons 
in the 126-184 shell, not shown in Table 1. In the latter case, the proton magic number 114
should not be used, since the recent production of nuclides up to A=118 has rendered it obsolete.  

The contents of the present Table 1 can be correlated to the contents of Table I of Ref. \cite{Cakirli},
where the orbit combinations for which $\delta V_{pn}$ is maximum are highlighted. The following remarks apply.

1) In $^{168}_{68}$Er$_{100}$, one has 18 valence protons and 18 valence neutrons. From Table 1 it is clear that 
the lowest 9 proton orbitals and the lowest 9 neutron orbitals are occupied, all of them forming 0[110] pairs.
The last pair, (p:7/2[523], n:7/2[633]), is indeed the 9th pair appearing in the rhs part of Table 1.

2) In $^{172}_{70}$Yb$_{102}$, one has 20 valence protons and 20 valence neutrons. From Table 1 it is clear that 
the lowest 10 proton orbitals and the lowest 10 neutron orbitals are occupied, all of them forming 0[110] pairs.
The last pair, (p:1/2[411], n:1/2[521]), is indeed the 10th pair appearing in the rhs part of Table 1.

3) In $^{176}_{72}$Hf$_{104}$, one has 22 valence protons and 22 valence neutrons. From Table 1 it is clear that 
the lowest 10 proton orbitals and the lowest 10 neutron orbitals are occupied, all of them forming 0[110] pairs.
In addition, the last protons occupy the 7/2[404] orbital, while the last neutrons occupy the 5/2[512] orbital,
both of them found on the 11th line of the rhs part of Table 1. This last pair is NOT of the 0[110] type, 
thus no maximum $\delta V_{pn}$ appears for this nucleus. The maximum appears for $^{178}_{72}$Hf$_{106}$,
in which the neutron orbital 7/2[514] is filled, meeting its proton 0[110] partner 7/2[404], which is already occupied.

4) In $^{180}_{74}$W$_{106}$, one has 24 valence protons and 24 valence neutrons. From Table 1 it is clear that 
the lowest 12 proton orbitals and the lowest 12 neutron orbitals are occupied, all of them forming 0[110] pairs,
thus maximum $\delta V_{pn}$ appears for this nucleus. 

The contents of the present Table 1 can also be correlated to the contents of Fig. 2.4 of Ref. \cite{IA}. 
The following remarks apply.

1) Well deformed nuclei, exhibiting the IBA SU(3) symmetry \cite{IA}, are expected to occur in the rare earth region
(p:50-82, n:82-126), up to 24 valence protons and 24 valence neutrons (i.e., up to $^{180}_{74}$W$_{106}$), 
since up to that point one can have purely 0[110] pairs. Beyond this point, the $n_z=0$ orbitals start to play a role, 
thus diluting maximum deformation. Near the end of the region, it is a reasonable approximation to use 
proton holes and neutron holes. From Table 1 it is clear that the relevant orbitals do NOT form 0[110] pairs. 
This is a region in which O(6) nuclei are known to occur, the textbook example being $^{196}_{78}$Pt$_{118}$. 
In this nucleus, proton holes are expected to occupy the 1/2[400] and 3/2[402] orbitals, while neutron holes are expected 
to occupy the 1/2[501], 5/2[503], 3/2[501], and 13/2[606] orbitals. Indeed, no 0[110] pairs occur. 

2) Well deformed nuclei, exhibiting the IBA SU(3) symmetry, are also expected to occur in the known part of the actinide region,
since one can see that $n_z=0$ orbitals appear beyond 36 valence neutrons. 

3) An O(6) region is known to occur for protons in the lower part of the 50-82 major shell
and neutrons in the upper part of the 50-82 shell, around A=130. In $^{130}_{54}$Xe$_{76}$, for example, 
one expects valence protons to occupy the 1/2[431] and 3/2[422] orbitals, with neutron holes occupying the 
11/2[505], 3/2[402], and 9/2[514] orbitals, while in $^{130}_{56}$Ba$_{74}$, for example, 
one expects valence protons to occupy the 1/2[431], 3/2[422], and 1/2[420] orbitals, with neutron holes occupying the 
11/2[505], 3/2[402], 9/2[514], and 1/2[400] orbitals. In both cases, no 0[110] pairs appear.  

4) Another O(6) region is known to occur for protons in the lower to middle part of the 28-50 major shell
and neutrons in the upper part of the 28-50 shell. Again, no 0[110] pairs occur. 

\section{Approximate SU(3) symmetry}

It is well known that the SU(3) symmetry of the 3D harmonic oscillator, appearing in light nuclei 
up to the sd shell \cite{Elliott}, is destroyed in heavier nuclei because of the strong spin-orbit interaction \cite{Mayer}.
This can be seen in Table 2. 


\begin{table} [htb]
\caption{Nilsson orbitals appearing in different major shells of neutrons (n) and protons (p) 
of the nuclear shell model and in different shells of the 3D harmonic oscillator.}\smallskip
\begin{small}\centering
\begin{tabular*}{\textwidth}{@{\extracolsep{\fill}}rrrr|rrrr}
\hline  \noalign {\smallskip}
28-50 & 28-50 &  pf  & pf  & 50-82 & 50-82 & sdg & sdg \\ 
n, p    & n, p   & n, p & n, p& n     &  n    &  n  &  n  \\ 
\hline\noalign {\smallskip}
2p1/2 & 1/2[301] & 2p1/2 & 1/2[301] & 3s1/2 & 1/2[411] & 3s1/2 & 1/2[411] \\
2p3/2 & 1/2[321] & 2p3/2 & 1/2[321] & 2d3/2 & 1/2[400] & 2d3/2 & 1/2[400] \\
      & 3/2[312] &       & 3/2[312] &       & 3/2[402] &       & 3/2[402] \\
1f5/2 & 1/2[310] & 1f5/2 & 1/2[310] & 2d5/2 & 1/2[431] & 2d5/2 & 1/2[431] \\
      & 3/2[301] &       & 3/2[301] &       & 3/2[422] &       & 3/2[422] \\
      & 5/2[303] &       & 5/2[303] &       & 5/2[413] &       & 5/2[413] \\
1g9/2 & 1/2[440] & 1f7/2 & 1/2[330] & 1g7/2 & 1/2[420] & 1g7/2 & 1/2[420] \\
      & 3/2[431] &       & 3/2[321] &       & 3/2[411] &       & 3/2[411] \\
      & 5/2[422] &       & 5/2[312] &       & 5/2[402] &       & 5/2[402] \\ 
      & 7/2[413] &       & 7/2[303] &       & 7/2[404] &       & 7/2[404] \\ 
      & 9/2[404] &       &          &1h11/2 & 1/2[550] & 1g9/2 & 1/2[440] \\
      &          &       &          &       & 3/2[541] &       & 3/2[431] \\ 
      &          &       &          &       & 5/2[532] &       & 5/2[422] \\ 
      &          &       &          &       & 7/2[523] &       & 7/2[413] \\ 
      &          &       &          &       & 9/2[514] &       & 9/2[404] \\  
      &          &       &          &       &11/2[505] &       &          \\ 
 \hline  \noalign {\smallskip}
50-82 & 50-82 & sdg & sdg & 82-126   & 82-126  &  pfh  & pfh   \\ 
 p     &  p    &  p  &  p  & n        &  n      &   n   &  n    \\
\hline\noalign{\smallskip}
3s1/2 & 1/2[400] & 3s1/2 & 1/2[400] & 3p1/2 & 1/2[501] & 3p1/2 & 1/2[501]  \\ 
2d3/2 & 1/2[411] & 2d3/2 & 1/2[411] & 3p3/2 & 1/2[521] & 3p3/2 & 1/2[521]  \\
      & 3/2[402] &       & 3/2[402] &       & 3/2[512] &       & 3/2[512]  \\
2d5/2 & 1/2[420] & 2d5/2 & 1/2[420] & 2f5/2 & 1/2[510] & 3f5/2 & 1/2[510]  \\
      & 3/2[411] &       & 3/2[411] &       & 3/2[501] &       & 3/2[501]  \\
      & 5/2[402] &       & 5/2[402] &       & 5/2[503] &       & 5/2[503]  \\
1g7/2 & 1/2[431] & 1g7/2 & 1/2[431] & 2f7/2 & 1/2[541] & 3f7/2 & 1/2[541]  \\
      & 3/2[422] &       & 3/2[422] &       & 3/2[532] &       & 3/2[532]  \\
      & 5/2[413] &       & 5/2[413] &       & 5/2[523] &       & 5/2[523]  \\
      & 7/2[404] &       & 7/2[404] &       & 7/2[514] &       & 7/2[514]  \\
1h11/2& 1/2[550] & 1g9/2 & 1/2[440] & 1h9/2 & 1/2[530] & 1h9/2 & 1/2[530]  \\
      & 3/2[541] &       & 3/2[431] &       & 3/2[521] &       & 3/2[521]  \\
      & 5/2[532] &       & 5/2[422] &       & 5/2[512] &       & 5/2[512]  \\
      & 7/2[523] &       & 7/2[413] &       & 7/2[503] &       & 7/2[503]  \\
      & 9/2[514] &       & 9/2[404] &       & 9/2[505] &       & 9/2[505]  \\
      &11/2[505] &       &          & 1i13/2& 1/2[660] &1h11/2 & 1/2[550]  \\
      &          &       &          &       & 3/2[651] &       & 3/2[541]  \\
      &          &       &          &       & 5/2[642] &       & 5/2[532]  \\
      &          &       &          &       & 7/2[633] &       & 7/2[523]  \\
      &          &       &          &       & 9/2[624] &       & 9/2[514]  \\
      &          &       &          &       &11/2[615] &       &11/2[505]  \\
      &          &       &          &       &13/2[606] &       &           \\

\hline
\end{tabular*}
\end{small}
\end{table} 

In the upper left part of Table 2, it is seen that the pf shell is losing the 1f7/2 orbital, which 
is lowered by the spin orbit interaction, gets isolated, and forms by itself the 20-28 major shell, 
while it gains the 1g9/2 orbital, coming down from the sdg shell. As a result, the 28-50 major shell is formed. 
Listing explicitly the Nilsson orbitals occuring in each case, and comparing the 28-50 and pf columns,
we see that the 28-50 shell can be treated approximately as a ``pf'' shell plus the single orbital 9/2[404]
(the one with highest $K$ in this shell).  This approximation is reasonable because the ``missing''
1/2[330], 3/2[321], 5/2[312], 7/2[303] orbitals are replaced by their 0[110] counterparts,
1/2[440], 3/2[431], 5/2[422], 7/2[413], which have the same orbital angular momentum projections
and total angular momentum projections, differing  only by one quantum of excitation along the $z$-axis. 
The approximation is also reasonable because the single orbital to be kept aside, 9/2[404],
lies highest in energy, as seen in Table 1. 

In the other parts of Table 2 we see that a similar approximation can also be made in the other major shells.

1) The 50-82 shell can be considered as a ``sdg'' shell, plus the single orbital 11/2[505]. 
The missing 1g9/2 levels are replaced by their 0[110] partner levels of  1h11/2. The single orbital
11/2[505] again lies high in energy, being the highest in the case of neutrons, and the third from the top 
of the shell in the case of protons.

2) The 82-126 shell can be considered as a ``pfh'' shell, plus the single orbital 13/2[606]. 
The missing 1h11/2 levels are replaced by their 0[110] partner levels of  1i13/2. The single orbital
13/2[606] again lies high in energy, being the 4th (out of 22) from the top in the case of neutrons, and the 7th (out of 22)
from the top of the shell in the case of protons.

3) In a similar way, the 126-184 shell can be considered as a ``sdgi'' shell, plus the single orbital 15/2[707]. 
The missing 1i13/2 levels are replaced by their 0[110] partner levels of  1j15/2. The single orbital
15/2[707] again lies high in energy, being the 6th (out of 29) from the top in the case of neutrons.

\section{Comparison to earlier work}

\subsection{The Federman--Pittel mechanism}

In the late '70ies, the Federman--Pittel mechanism has been introduced \cite{FP1,FP2}, stating that
the $^3$S$_1$ component of the n-p force between 
spin-orbit pairs of protons and neutrons plays the major role in the creation of nuclear deformation.

In the $_{40}$Zr-$_{42}$Mo region, considering the isotopes up to N=62 and 64 respectively,
it has been argued \cite{FP1} that the 1g9/2 proton and 1g7/2 neutron orbitals play the major role 
in creating deformation. A later Hartree--Fock--Bogolyubov study \cite{FP2} indicated that this is true in the 
beginning of the shell, while later on the 1g9/2 proton and 1h11/2 neutrons play the major role 
in creating deformation. This is in qualitative agreement with the present approach, since the 
1g9/2 and 1h11/2 orbitals are made up by 0[110] partners, as seen in Table 2. Moreover, these 
orbitals start being filled around the proton midshell and quite early in the neutron shell, 
as seen in Table 1.        

Concerning the rare earth region, it was conjectured \cite{FP1} that the proton 1h11/2 and neutron 1h9/2 
orbitals are important in the beginning of the region, while later on the proton 1h11/2 and neutron 1i13/2 orbitals
play the major role, until the end of the deformed rare-earth region around Z$\sim 76$ and N$\sim 114$.
The latter is in agreement with the present approach, since the 1h11/2 and 1i13/2 orbitals are made from 
0[110] partners, as seen in Table 2.
 
Concerning the actinide region, it was conjectured \cite{FP1} that the proton 1i13/2 and neutron 1i11/2 
orbitals are important in the beginning of the region, while later on the proton 1i13/2 and neutron 1j15/2 orbitals
play the major role. The latter is in agreement with the present approach, since the 1i13/2 and 1j15/2 orbitals are made from 
0[110] partners. 

It should be noticed that the other pairs suggested by Federman--Pittel are also related in a simple way, as described below.

1) In the $_{40}$Zr-$_{42}$Mo region, the 1g9/2 proton and 1g7/2 neutron orbitals consist of three 0[020] pairs plus one 0[01$\bar 1$] pair
(using the notation $\bar 1$ for $-1$), as seen in Table 2.

2) In the rare earth region, the proton 1h11/2 and neutron 1h9/2 orbitals consist of four 0[020] pairs plus one 0[01$\bar 1$] pair, 
as seen in Table 2.

2) In the actinides, the proton 1i13/2 and neutron 1i11/2 orbitals consist of five 0[020] pairs plus one 0[01$\bar 1$] pair.

In Refs. \cite{FP1,FP2} it is argued that the orbitals 1d5/2 and 1d3/2 play a major role in the development 
of deformation in the sd shell. One can see that these orbitals can form one 0[020] pair plus one 0[01$\bar 1$] pair.   

\subsection{The pseudo-SU(3) scheme}

In the early '70ies the pseudo-SU(3) scheme \cite{pseudo1,pseudo2} has been proposed, 
in an effort to extend the Elliott SU(3) symmetry beyond the sd shell. 

In the pseudo-SU(3) scheme, in a given major shell with quantum number N, the orbitals remaining after the departure 
of the highest $K$ orbital, which jumps into the shell below, are assigned pseudo-SU(3) 
quantum numbers corresponding to a shell with $\tilde N=N-1$, while the opposite parity orbital, 
coming from the shell above, is kept separate. In this way 

1) In the 28-50 shell, the orbitals 2p1/2, 2p3/2, 1f5/2, with $N=3$ are assigned the pseudo-SU(3)
labels $\tilde s$1/2, $\tilde d$3/2, $\tilde d$5/2, in a shell with $\tilde N=2$,
while the opposite parity 1g9/2 orbital (accommodating 10 particles) is kept separate. 

2) In the 50-82 shell, the orbitals 3s1/2, 2d3/2, 2d5/2, 1g7/2 with $N=4$ are assigned the pseudo-SU(3) labels 
$\tilde p$1/2, $\tilde p$3/2, $\tilde f$5/2, $\tilde f$7/2 in a shell with $\tilde N=3$, while the 
opposite parity 1h11/2 orbital (accommodating 12 particles) is kept separate.      

3) In the 82-126 shell, the orbitals 3p1/2, 3p3/2, 2f5/2, 2f7/2, 1h9/2 with $N=5$ are assigned the pseudo-SU(3)
labels $\tilde s$1/2, $\tilde d$3/2, $\tilde d$5/2, $\tilde g$7/2, $\tilde g$9/2 in a shell with $\tilde N=4$,
while the opposite parity 1i13/2 orbital (accommodating 14 particles) is kept separate. 

4) In the 126-184 shell, the orbitals 4s1/2, 3d3/2, 3d5/2, 2g7/2, 2g9/2, 1i11/2, with $N=6$ are assigned the pseudo-SU(3) labels 
$\tilde p$1/2, $\tilde p$3/2, $\tilde f$5/2, $\tilde f$7/2, $\tilde h$9/2, $\tilde h$11/2, in a shell with $\tilde N=5$, while the 
opposite parity 1j15/2 (accommodating 16 particles) orbital is kept separate.      

The difference between the pseudo-SU(3) scheme and the present approach is clear. In the pseudo-SU(3) scheme, in a given major shell,
an approximate SU(3) is arising from the orbitals having the same parity, while the single orbital having the opposite parity 
remains isolated. In the present scheme, orbitals of both parities are taken into account on equal footing 
in creating an approximate SU(3), with only one Nilsson orbital (accommodating two particles) remaining isolated.    

\subsection{Detailed consideration of the pseudo-SU(3) scheme}

Let us consider the proton 50--82 major shell. The Nilsson orbitals with positive parity 
appearing in this shell are listed in Table 3. The pseudo-SU(3) orbitals onto which they are mapped, 
as described in detail in Fig. 1 of Ref. \cite{pseudo1}, are also shown in Table 3.
The following remarks apply.


\begin{table} [htb]
\caption{Mapping of the negative parity orbitals of the 28--50 major shell, 
the positive parity orbitals of the 50--82 major shell,
and of the negative parity orbitals of the 82--126 shell
 onto pseudo-SU(3) shells.}\smallskip
\begin{small}\centering
\begin{tabular*}{\textwidth}{@{\extracolsep{\fill}}rr|rr|rr}
\hline  \noalign {\smallskip}
28-50 & pseudo sd & 50-82  & pseudo pf  &  82-126 & pseudo sdg \\ 
\hline\noalign
{\smallskip}
 1/2[321] & 1/2[220] & 1/2[431] & 1/2[330] & 1/2[541] & 1/2[440] \\
 1/2[310] & 1/2[211] & 1/2[420] & 1/2[321] & 1/2[530] & 1/2[431] \\
 3/2[312] & 3/2[211] & 3/2[422] & 3/2[321] & 3/2[532] & 3/2[431] \\
 1/2[301] & 1/2[200] & 3/2[411] & 3/2[312] & 3/2[521] & 3/2[422] \\
 3/2[301] & 3/2[202] & 5/2[413] & 5/2[312] & 5/2[523] & 5/2[422] \\
 5/2[303] & 5/2[202] & 1/2[411] & 1/2[310] & 1/2[521] & 1/2[420] \\
          &          & 5/2[402] & 5/2[303] & 5/2[512] & 5/2[413] \\
          &          & 7/2[404] & 7/2[303] & 7/2[514] & 7/2[413] \\
          &          & 1/2[400] & 1/2[301] & 7/2[503] & 7/2[404] \\
          &          & 3/2[402] & 3/2[301] & 9/2[505] & 9/2[404] \\
          &          &          &          & 1/2[510] & 1/2[411] \\
          &          &          &          & 3/2[512] & 3/2[411] \\
          &          &          &          & 3/2[501] & 3/2[402] \\
          &          &          &          & 5/2[503] & 5/2[402] \\
          &          &          &          & 1/2[501] & 1/2[400] \\ 
\hline
\end{tabular*}
\end{small}
\end{table} 
          
1) Orbitals with $N=4$ are mapped onto orbitals with $\tilde N=3$. 
In other words, $N$ is changed and parity is changed.

2) In all cases $\Lambda$ is changed by one unit, either positive or negative.
Thus the projection of orbital angular momentum is modified. 

3) In all cases, spin is inverted. For example, 1/2[431] has spin down, while 
its pseudo-SU(3) image, $1/2\tilde{[330]}$, has spin up. 
Thus the projection of spin is modified. 

4) However, the changes in the projections of orbital angular momentum and of spin, mentioned 
in 2) and 3), are done in such a way that the projection of the total angular momentum,
$\Omega$, remains intact.     

5) In this way, the positive parity orbitals of the 50--82 proton shell, which come from 
an sdg shell from which 1g9/2 has escaped, thus breaking the U(15) symmetry, 
are mapped onto a complete pseudo-pf shell with U(10) symmetry. 

All the above regard the 10 orbitals of the 50--82 shell having positive parity.
The 6 orbitals having negative parity are left aside. They belong to 1h11/2, having a U(12) symmetry.

The neutron 82--126 major shell is also shown in Table 3. The mapping is performed as desribed in 
Fig. 2 of Ref. \cite{pseudo1}. The following comments apply.

1) Orbitals with $N=5$ are mapped onto orbitals with $\tilde N=4$.

2) Comments 2), 3), 4) listed above, are valid unchanged.  
  
3) In this way, the negative parity orbitals of the 82--126 neutron shell, which come from 
a pfh shell from which 1h11/2 has escaped, thus breaking the U(21) symmetry, are mapped onto a complete pseudo-sdg shell with U(15) symmetry.   
  
All the above regard the 15 orbitals of the 82--126 shell having negative parity.
The 7 orbitals having positive parity are left aside. They belong to 1i13/2, having a U(14) symmetry.   
  
\subsection{Comparison to the new coupling scheme} 

The handling of the 50-82 proton shell in the new coupling scheme can be seen in Table 2
(lower left corner). The 10 positive parity orbitals are mapped onto themseleves, 
i.e., they remain intact. Out of the 6 orbitals with negative parity, belonging to 1h11/2, the first 5 are mapped onto 
an 1g9/2 shell, while the last one, 11/2[505], remains isolated. The mapping of the 5 negative parity orbitals 
bears the following features.

1)  Orbitals with $N=5$ are mapped onto orbitals with $N=4$.
In other words, $N$ is changed and parity is changed. This change is similar 
to the one occuring in pseudo-SU(3) for the positive parity orbitals. 

2) The projections of orbital angular momentum $\Lambda$ and spin $\Sigma$ remain intact, 
while in the pseudo-SU(3) scheme they are changed.

3) The projection of the total angular momentum $\Omega$ remains intact, as in the pseudo-SU(3) scheme. 
  
4) The original orbitals and their images are related by 0[110], i.e., 
$n_z$ is also changed by one unit.   
  
5) In this way, the orbitals of the 50--82 proton shell, which come from 
an sdg shell from which 1g9/2 has escaped, thus breaking the U(15) symmetry,
and to which the negative parity 1h11/2 has been added in order to make the situation worse, 
are mapped onto a complete sdg shell with U(15) symmetry, by just leaving aside the 11/2[505] orbital. 
In the Nilsson diagrams one can see that this orbital starts at high energy within the shell and for prolate 
deformations its slope is strongly upwards. Thus it is the highest lying orbital for large deformations,
while it is among the 3 highest lying orbitals even at small deformations. 

Let us now focus on the handling of the 82-126 neutron shell in the new coupling scheme, which can be seen in Table 2
(lower right corner). The 15 negative parity orbitals are mapped onto themseleves, 
i.e., they remain intact. Out of the 7 orbitals with positive parity, belonging to 1i13/2, the first 6 are mapped onto 
an 1h11/2 shell, while the last one, 13/2[606], remains isolated. The mapping of the 6 positive parity orbitals 
bears the following features.

1)  Orbitals with $N=6$ are mapped onto orbitals with $N=5$.
In other words, $N$ is changed and parity is changed. This change is similar 
to the one occuring in pseudo-SU(3) for the negative parity orbitals. 

2) The projections of orbital angular momentum $\Lambda$ and spin $\Sigma$ remain intact, 
while in the pseudo-SU(3) scheme they are changed.

3) The projection of the total angular momentum $\Omega$ remains intact, as in the pseudo-SU(3) scheme. 
  
4) The original orbitals and their images are related by 0[110], i.e., 
$n_z$ is also changed by one unit.   
  
5) In this way, the orbitals of the 82--126 neutron shell, which come from 
a pfh shell from which 1h11/2 has escaped, thus breaking the U(21) symmetry,
and to which the positive parity 1i13/2 has been added in order to make the situation worse, 
are mapped onto a complete pfh shell with U(21) symmetry, by just leaving aside the 13/2[606] orbital. 
In the Nilsson diagrams one can see that this orbital starts at high energy within the shell and for prolate 
deformations its slope is strongly upwards. Thus it is the highest lying orbital for large deformations,
while it is among the 3 highest lying orbitals even at small deformations. 
 
Therefore a comparison between the pseudo-SU(3) scheme and the present coupling scheme boils down 
to the following points.

1) In the pseudo-SU(3) scheme only the normal parity levels are mapped, while the opposite parity levels 
are left intact. In this way a pseudo-SU(3) symmetry for the normal parity levels is obtained, 
with $N$ being lowered by one unit. In the new coupling scheme, the normal parity levels are kept intact, 
with only the opposite parity levels being mapped (except one) onto levels with $N$ lower by one unit.
In this way the original SU(3) symmetry of the shell under discussion is restored, while one high-lying orbital 
(being able to accommodate two particles) remains isolated. 

2) The pseudo-SU(3) scheme is based on the similar behavior of pairs of orbitals, called spin--orbit partners,
within the same major shell, as explained in detail in Ref. \cite{pseudo1}. One can see that in all cases these are orbitals 
differing by 1[002]. In the proton diagram for the 50--82 shell,
for example, the orbitals 3/2[422] and 1/2[420] are represented by two lines being close to each other at all deformations. 
In contrast, the new coupling scheme is based on the similar behavior of pairs of orbitals, called 0[110] partners,
within two different but adjacent major shells. For example, the 1/2[550] orbital within the 50--82 major shell and the
1/2[660] orbital within the 82--126 major shell exhibit the same behavior with increasing deformation.  
Pseudo-SU(3) takes advantage of intra-shell similar behaviors, while the new coupling scheme takes advantage of inter-shell
similar behaviors.    
 
\section{Implications for microscopic calculations}

The pseudo-SU(3) scheme has been implemented in microscopic calculations 
for heavy deformed nuclei. In Refs. \cite{DW1,DW2}, for example, a detailed study 
for 8 deformed rare earths and 4 deformed actinides has been carried out.
It should be interesting to see how the results of this study are modified 
if the present approximate SU(3) scheme is employed. Some preliminary results 
have been reported in Ref. \cite{Bonatsos}, while further work is in progress. 

\section*{Acknowledgments}

Financial support from the Bulgarian National Science Fund under 
Contract No. DFNI-E02/6 is gratefully acknowledged.

\end{document}